\documentclass{article}
\usepackage{amssymb}
\usepackage{epsfig}
\usepackage{rotating}
\usepackage{floatflt}
\usepackage{cite}
\usepackage{here}
\newcommand{\bc}{\begin{center}}
\newcommand{\ec}{\end{center}}

\begin{document}
\bc{\Large \bf Renormalization scheme dependence in the case of a QCD
non-power perturbative expansion}\ec
\bc{\large Jan Fischer, Ji\v{r}\'{\i} Ch\'{y}la}\ec
\bc{Institute of Physics, Academy of Sciences of the Czech Republic\\
Na Slovance 2, 18221 Prague 8, Czech Republic}\ec
\bc and \ec
\bc{\large Irinel Caprini}\ec
\bc {National Institute of Physics and Nuclear Engineering,
 POB MG6, R-76900 Bucharest,  Romania}\ec

\bc{\large Abstract}\ec
 A novel, non-power, expansion of QCD quantities replacing the
standard perturbative expansion in powers of the renormalized
couplant $a$ has recently been introduced and examined by two of
us. Being obtained by analytic continuation in the Borel plane, the new
expansion functions $W_{n}(a)$  share the basic analyticity
properties with the expanded quantity. In this note we
investigate the renormalization scale dependence of finite order
sums of this new expansion for the phenomenologically interesting
case of the $\tau$-lepton decay rate.
%\pacs{11.15.Bt,12.38.Aw,12.38.Cy,13.35.Dx}

\section{Renormalization scale and scheme dependence}
\label{intro}
        %%%%%%%%%%%%%%%%%%%%%%%%%%%%%%%%%%%
In the standard perturbation theory the finite order
approximations of physical quantities  are renormalization scale
($\mu$) and scheme (RS) dependent.
%(the adjective ``renormalization'' will be dropped in the following).
The quest
for in some sense ``optimal'' scale and scheme is vital for
meaningful applications  but has so far no generally accepted
solution. There are several recipes \cite{PMS,FAC,BLM}
how to do that.
The one proposed in \cite{PMS} and known as the Principle of
Minimal Sensitivity (PMS) selects the scale and scheme  by the
condition of local scale and scheme invariance. For a physical
quantity ${\cal R}(Q)$ depending on one external kinematical variable
$Q$ and admitting the perturbative expansion of the form
\begin{equation}
{\cal R}(Q)=a(\mu, {\rm RS})(1 + r_1(Q,\mu,{\rm RS})a(\mu,{\rm RS})
+ r_{2}(Q,\mu,{\rm RS}) a^{2}(\mu,{\rm RS}) + . . . ),~~~~~
a\equiv \alpha_{s}/\pi,
\label{pert}
\end{equation}
this implies for the sum $R^{(N)}(Q,\mu,{\mathrm{RS}})$ of first $N$ terms
in (\ref{pert})
\begin{equation}
{{\partial \cal{R}}^{(N)}(Q,\mu, {\rm RS})\over {\partial \ln \mu}}
\bigg|_{opt}= \frac{{\partial \cal{R}}^{(N)}(Q,\mu, {\rm RS)}}{\partial ({\rm
RS})} \bigg|_{opt} = 0 . \label{opt}
\end{equation}
The PMS thus selects the point where the truncated approximant has
locally the property  which the all order sum must have globally.
In the absence of additional information this choice appears
particularly well-motivated. But even if we do not subscribe to
PMS it is definitely useful to investigate the scale and scheme
dependence of finite order approximants. The scheme can be
labeled by the set of free parameters $c_k, k\ge 2$, defining the
r.h.s. of the RG equation for the couplant $a$
\begin{equation} {{\partial a}(\mu, {\rm RS})\over {\partial \ln \mu}}
 = \beta (a)
= -b a^{2} (1 + c a + c_{2}a^{2} + c_{3} a^{3} +\cdots),
\label{beta}
\end{equation}
together with some parameter specifying which of the solution of
eq. (\ref{beta}) we have in mind.   %We adopt the parameter $\tilde \Lambda$
One way of doing this is by means of the parameter
defined by the condition $a(\mu=\tilde \Lambda)=\infty$.
Note that the first two coefficients in (\ref{beta}),
$b=(33-2 n_f)/6$,  $c=(153- 19 n_f)/(66-4 n_f)$, are universal, and
that ${\tilde  \Lambda}$ defined above is related to the more commonly
used definition of $\Lambda$ by a simple scale factor close to unity:
${\tilde \Lambda} = \Lambda (2c/b)^{-c/b}$.

At the second order there are two
free parameters: the scale $\mu$ and $\tilde \Lambda$, specifying
the scheme, but without loss of generality we can fix the latter
and vary the scale only
\begin{equation}
{\cal R}^{(2)}(Q,\mu) = a^{(2)}(\mu)[1+r_{1}(Q,\mu)a^{(2)}(\mu)],
\label{R2}
\end{equation}
where $a^{(2)}(\mu)$ solves (\ref{beta}) with the first two terms
on its r.h.s. only and satisfies
\begin{equation}
b \ln (\mu/\tilde\Lambda)=1/a^{(2)}(\mu)+c\ln[c
a^{(2)}(\mu)/(1+ca^{(2)}(\mu))]. \label{trunc}
\end{equation}
The formal (i.e. to the order considered) scale independence of
(\ref{R2}) implies
\begin{equation}
\partial r_1(Q,\mu)/\partial \ln \mu=b,~~~~ \Rightarrow ~~~~
 r_1(Q,\mu)=b \ln (\mu/\tilde
\Lambda)-\rho_1(Q/\tilde \Lambda)\, , \label{r1}
\end{equation}
where $\rho_1$ is a scale and scheme invariant depending on $Q$
and the numerical value of ${\tilde \Lambda}$, which can be evaluated
using the results in $\overline{\mathrm{MS}}$ RS as
$\rho_1=b\ln(Q/\tilde{\Lambda}_{\overline{\mathrm{MS}}})
-r_1(\mu=Q,\overline{\mathrm{MS}})$.

At the third order, the coefficients $r_2$ in (\ref{pert}) and
$c_2$ in (\ref{beta}) come into play. As a consequence, both $r_2$
and the couplant $a$ depend beside $\mu$ and RS also on $c_2$. We
refer to \cite{Chyla} for details and mention only the expression
for $r_2$ which will be used in the following
\begin{equation}
r_2=\rho_2-c_2+(r_1+c/2)^2\,, \label{r2}
\end{equation}
where $\rho_2$ is another scale and scheme invariant, which unlike
$\rho_1$, is a pure number. Although at the third order $c_2$
is a free parameter, we shall not exploit the associated freedom,
but will work in the RS where $c_2=0$ at all orders.
We prefer this choice of the RS to the conventional
$\overline{\mathrm{MS}}$ RS since in this case
the coupling $a(\mu)$ is well defined and the same at
all orders, and any manifestation of the divergence of
perturbation expansion concerns exclusively the coefficients of
the expansion (\ref{pert}).

%%%%%%%%%%%%%%%%%%%%%%%%%%%%%%%%%%%%%
\section{Non-power expansions}
%%%%%%%%%%%%%%%%%%%%%%%%%%
In \cite{CaFi1,CaFi2,CaFi3} a method was proposed that replaces
perturbative expansions of observables in powers
of the QCD couplant $a$ by expansions in the set of
functions $W_n(a)$ encompassing the available knowledge of
the large order behaviour of standard perturbative expansions.
As an example we consider the phenomenologically interesting observable
\begin{equation}
R_{\tau} = \frac{\Gamma(\tau\rightarrow \nu_{\tau}+{\mathrm {hadrons}})}
{\Gamma(\tau\rightarrow \nu_{\tau}+{\mathrm {e}}^-\overline{\nu}_{e})}
= 3(1+\delta_{\mathrm{EW}})(1 + {\cal R}_{\tau}),
\label{RR}
\end{equation}
where $\delta_{\mathrm{EW}}$ is an electroweak correction and the QCD
contribution
${\cal R}_{\tau}$ is of the form (\ref{pert}). As shown in
\cite{Alta}, the term ${\cal R}_{\tau}$
can be formally written in the form of the Borel transform
\begin{equation}
{\cal R}_{\tau}(M_{\tau})=\int_{\cal C}{\rm e}^{-u/a(\mu)} B(\mu,u) F(bu/2){\rm d}u
\label{Borelint}
\end{equation}
involving the functions
\footnote{Eqs. (\ref{Borelint},\ref{FB}) were derived
in \cite{Alta} using the one-loop expression for the analytic continuation of
$a(-s)$ from Euclidean to Minkowskian region in the formula relating $D(s)$ to
$R_{\tau}$, thus, setting $c=0$. Using the NLO expression (\ref{trunc})
for $a(-s)$ would lead to a more complicated relation between $B(u)$ and
$R_{\tau}$. However, as we use $F(u)$ merely to define our expansion functions,
we can use the expression derived in \cite{Alta}, still retaining a consisnent
expansion of $R_{\tau}$ in terms of our functions to all orders.}
\begin{equation}
F(u) = \frac{-12 \sin(\pi u)}{\pi u(u-1)(u-3)(u-4)},~~~~
B(\mu,u)=\sum_{n=0}^{\infty}\frac{D_{n+1}(M_{\tau},\mu)}{n!}u^n.
\label{FB}
\end{equation}
In (\ref{Borelint}) we have written explicitly the dependence on the
arbitrary scale $\mu$ but suppressed that on $M_{\tau}$.
The coefficients $D_n$ come from the perturbative expansion of
the Adler function $D_{\tau}(s)$ in the Euclidean region $s<0$
\begin{equation}
\!\!D_{\tau}(s)=
D_{\tau}^{(0)}\left(1+D_1(\kappa)a(\kappa\sqrt{-s})+D_2(\kappa)
a^2(\kappa\sqrt{-s})+
\cdots\right),~
D_{\tau}^{(0)}=3(1+\delta_{\mathrm{EW}}),
\label{Adler}
\end{equation}
where the scale ambiguity is now parameterized via the
parameter $\kappa$ relating $\mu$ to $s$:
$\mu=\kappa \sqrt{-s}$.
The contour $\cal C$ runs from $0$ to $\infty$, circumventing the
singularities of $B(u)$, which create
%. In the case of their presence, the ambiguity in
%the choice of $\cal C$ implies
non-uniqueness of the integral (\ref{Borelint}).
%as in \cite{CaFi2},
We choose the principal
value prescription.

Following \cite{CaFi1,CaFi2,CaFi3,CiFi} we expand $B(u)$ in powers of a special
function $w(u)$ that maps the holomorphy domain of $B(u)$
(or its known part) onto a unit circle. For $D$ and
${\cal R}_{\tau}$, $w(u)$ has the form
\begin{equation}
w(u)=\frac{\sqrt{1+u}-\sqrt{1-u/2}}{\sqrt{1+u}+\sqrt{1-u/2}},
\label{w}
\end{equation}
which enters the definition of the functions $W_n(a)$
\begin{equation}
W_{n}(a) \equiv \frac{1}{n!} \left (\frac{8}{3}\right)^n
\left(\frac{2}{b}\right)^n { 2\over a b} \int_{\cal C}
{\rm e}^{-2 u/(a  b)}F(u)
w^{n}(u) {\rm d}u
\label{Wn}
\end{equation}
relevant for
%the expansion of
${\cal R}_{\tau}$. For $D$ the $W_n(a)$ are also given by (\ref{Wn}),
but with $F(u)=1$. The $W_{n}(a)$ take into
account the positions $u = -2/b$, and
$u=4/b$ of the two leading singularities of $B(u)$.
But we also know that these singularities
%For $D_{\tau}$ and $\cal R_{\tau}$, we know also the {\it nature} of th
%the leading singularities
have the form $(1+ub/2)^{\gamma_1}$ and $(1-ub/4)^{\gamma_2}$
with $\gamma_1=-2.589$ and $\gamma_2=-2.58$. To use this knowledge
%of their {\it nature} (i.e., not only location)
we define
%the functions  that take this information into account
\begin{equation}
{\widetilde W}_{n}(a) = \frac{1}{n!} \left({8\over
3}\right)^n \left({2\over b}\right)^n \frac{2}{ab}
\int_{\cal C}{\rm e}^{-2 u/(a b)} (1+u)^{\gamma_1}
(1-u/2)^{\gamma_2} F(u) \, w^{n}(u) {\rm d}u
\label{WnFmod}
\end{equation}
and expand $R_{\tau}$ in terms of them. We explore the scale dependence of
both expansions, in $W_{n}$ and in ${\widetilde W}_{n}(a)$.
The incorporation of the nature of a singularity turns out to be quite important.
For details about the functions $W_n$
and $\widetilde W_n$ see \cite{CaFi1,CaFi2,CaFi3}.

As was shown in \cite{CaFi2}, expansions in terms of the $W_n$ or
$\widetilde{W}_n$ are convergent under rather loose conditions on the
coefficients. On the other hand, the functions themselves are singular at $a=0$
\cite{CaFi3}, the series
\begin{equation}\label{Wcorr}
W_n(a)\sim \sum\limits_{j\ge n} c_{nj} a^j,~~~~~
\widetilde{W}_n(a)\sim \sum\limits_{j\ge n} \tilde{c}_{nj} a^j
\label{sim}
\end{equation}
being asymptotic. We choose the normalization
such that $c_{nn}=1,\tilde{c}_{nn}=1,~ n\ge 1$.

%%%%%%%%%%%%%%%%%%%%%%%%%%%%
\section{Renormalization scale dependence for non-power expansions}
The scale and scheme dependence of ${\cal R}_{\tau}$ in the standard
perturbation theory was discussed in
\cite{Chyla,Alta}. In terms of the functions $W_n(a)$ or
$\widetilde W_n(a)$ we can rewrite ${\cal R}_{\tau}$ as
\begin{equation}
{\cal R}_{\tau}={\cal W}_1(a)+\overline{r}_1{\cal W}_2(a)+\overline{r}_2
{\cal W}_3(a)+\cdots,~~{\cal W}_n=W_n~~{\mathrm {or}}~\widetilde W_n,
\label{Wpert}
\end{equation}
where the coefficients $\overline{r}_k(M_{\tau},\mu)$ are related to
the $r_k(M_{\tau},\mu)$ of (\ref{pert}) as follows
\begin{equation}
\overline{r}_1=r_1-c_{12},
~~~~\overline{r}_2=r_2-\overline{r}_1 c_{23}-c_{13},
~~{\mathrm{etc.}}
\label{relations}
\end{equation}
The finite sums ${\cal R}_W^{(N)}$ of the first $N$ terms in the expansion
(\ref{Wpert}) have the same
property of formal scale independence as the conventional finite sums
in powers of the coupling $a$, i.e. their derivatives with
respect to $\ln \mu$ start at the order $N+1$
\begin{equation}
\frac{\partial {\cal R}_W^{(N)}(\mu)}{\partial \ln\mu}
=\sum_{\scriptstyle k=N+1}^{\infty}s_k W_k(a),
\label{cons}
\end{equation}
where $s_k$ are some numbers, which is a generalization
of the analogous relation in the conventional perturbation theory.
In our numerical studies we set $Q=M_{\tau}=1.8$ GeV in the expression for
the invariant $\rho_1(Q)$,  and
took $b=4.5,c=1.8,\rho_2=-6.27$, corresponding to $n_f=3$ \cite{Chyla}.
In the NLO we work in standard $\overline{\mathrm{MS}}$ scheme, in the
NNLO in the scheme where $c_2=0$. We did not resort to the
conventional practice of expanding the solution of eq. (\ref{trunc}) in
inverse powers of $\ln(\mu/\Lambda)$, but solved this equation numerically.

\begin{figure}\unitlength=1mm
\begin{picture}(160,180)
\put(0,130){\psfig{file=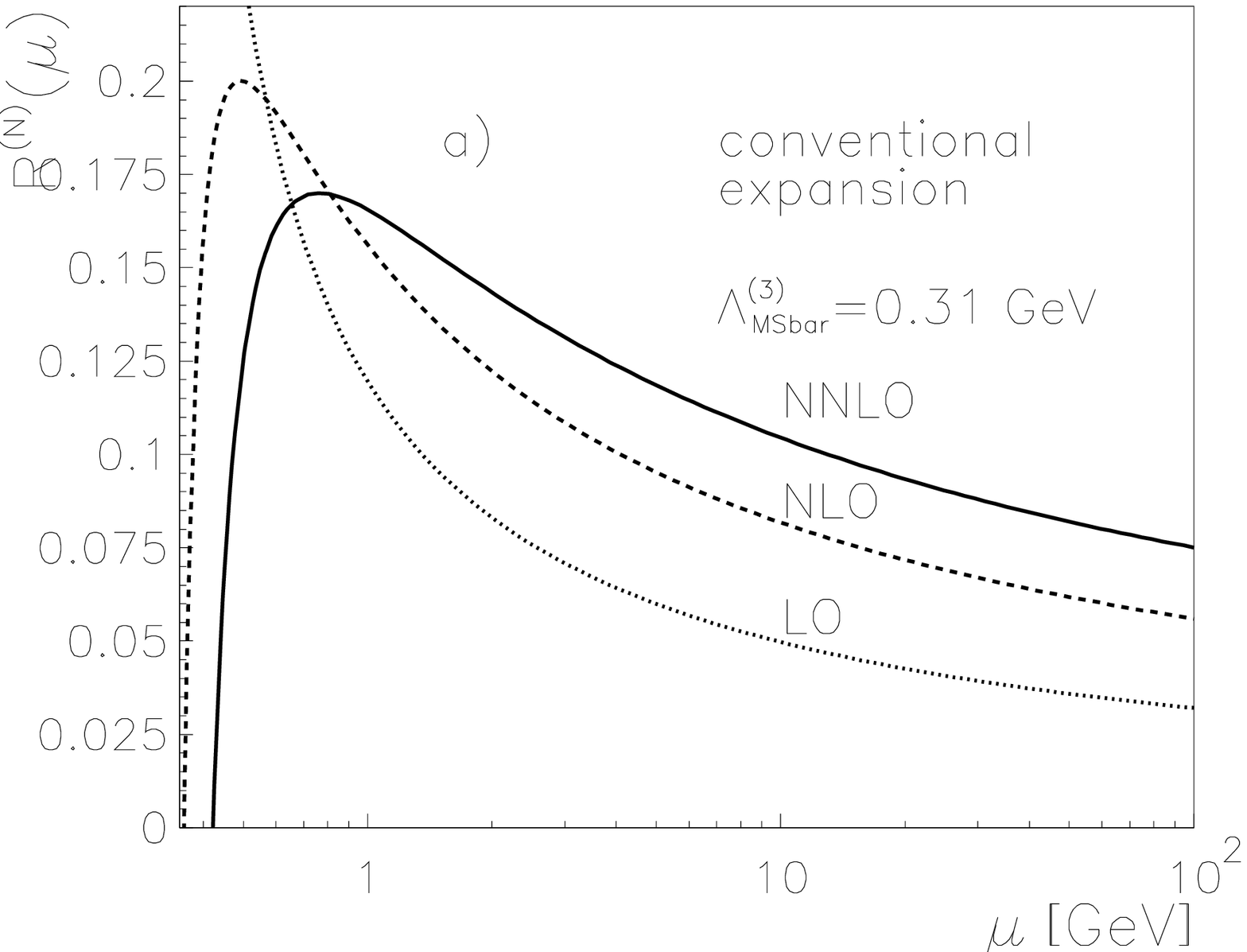,width=8cm}}
\put(80,130){\psfig{file=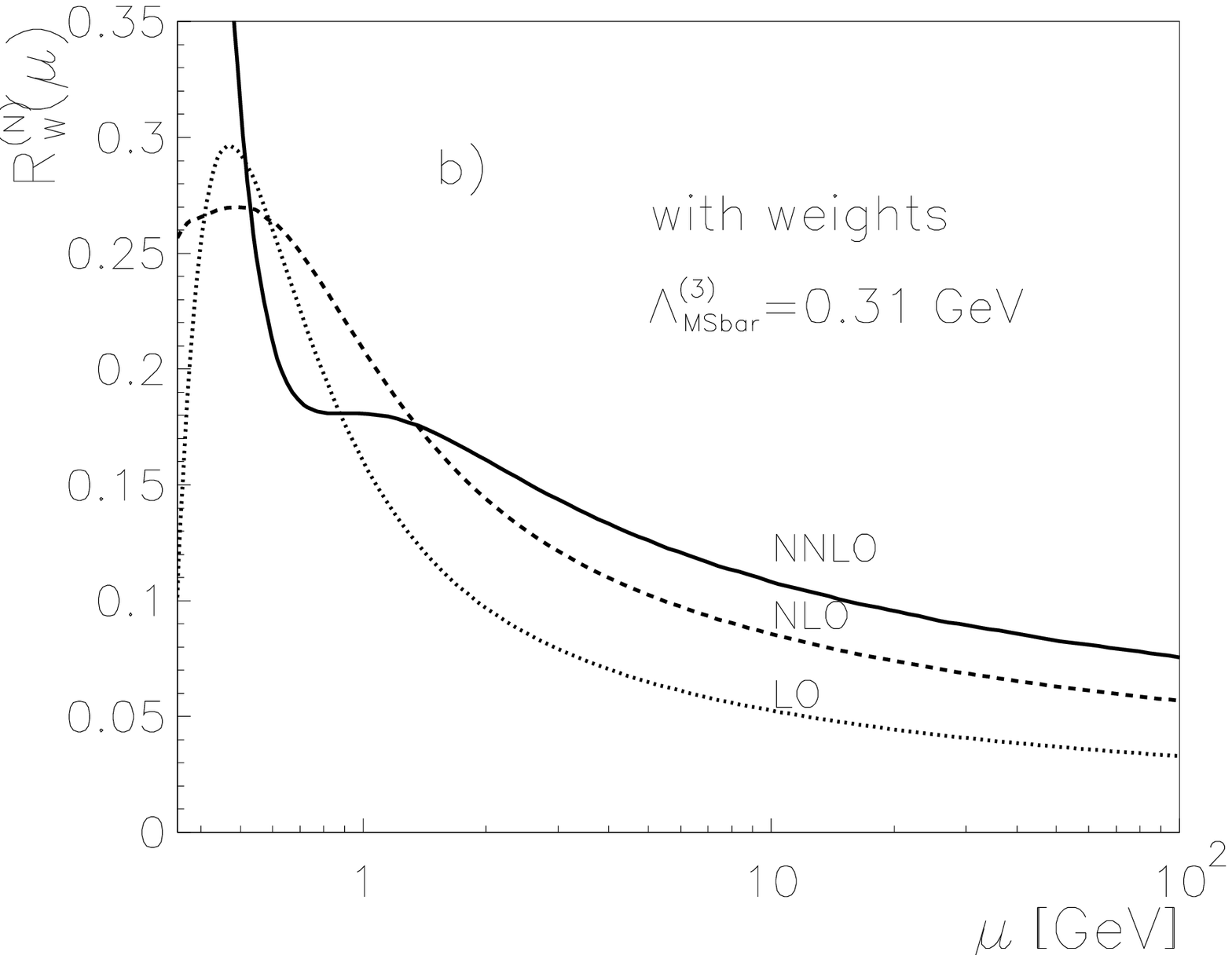,width=8cm}}
\put(80,65){\psfig{file=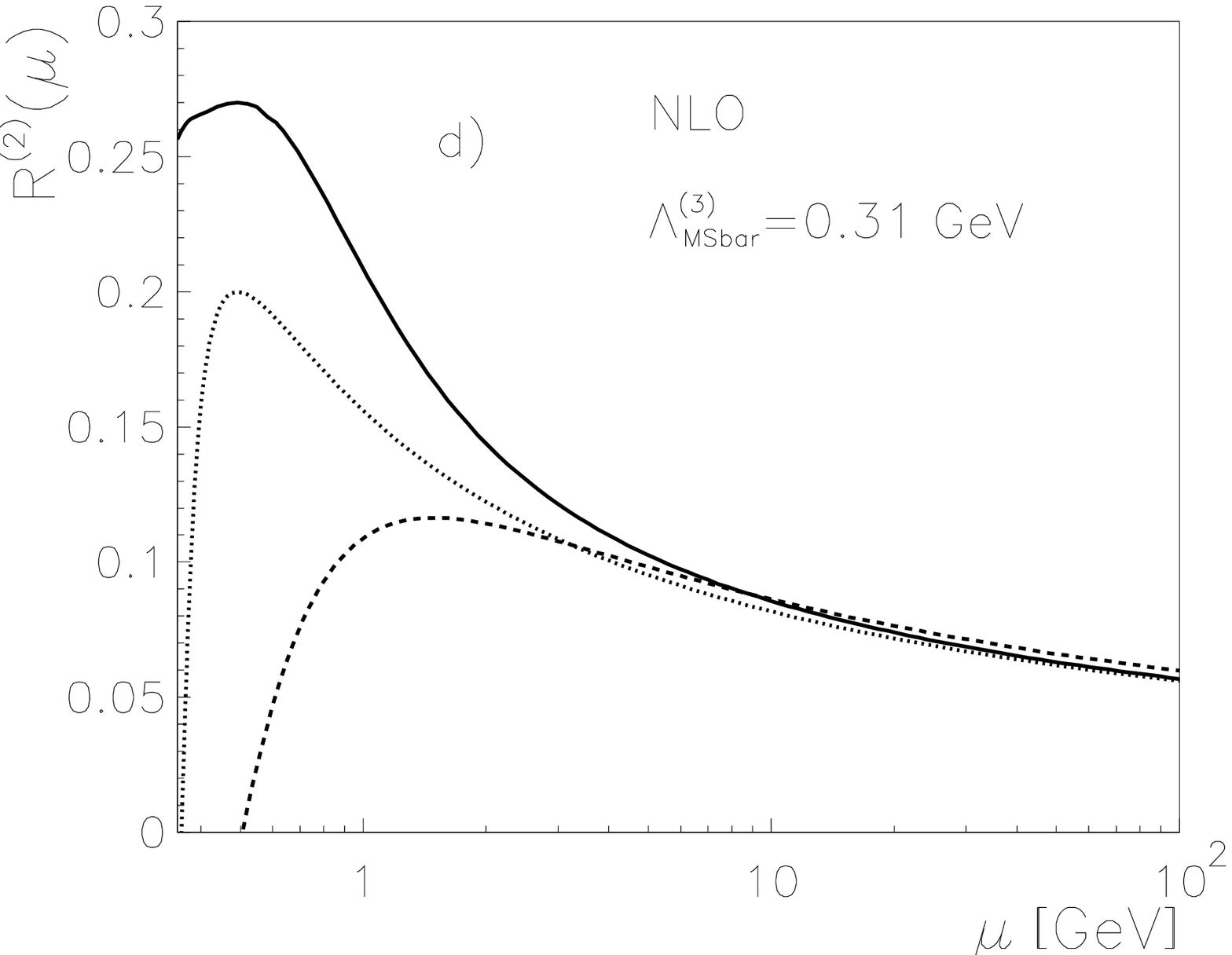,width=8cm}}
\put(0,0){\psfig{file=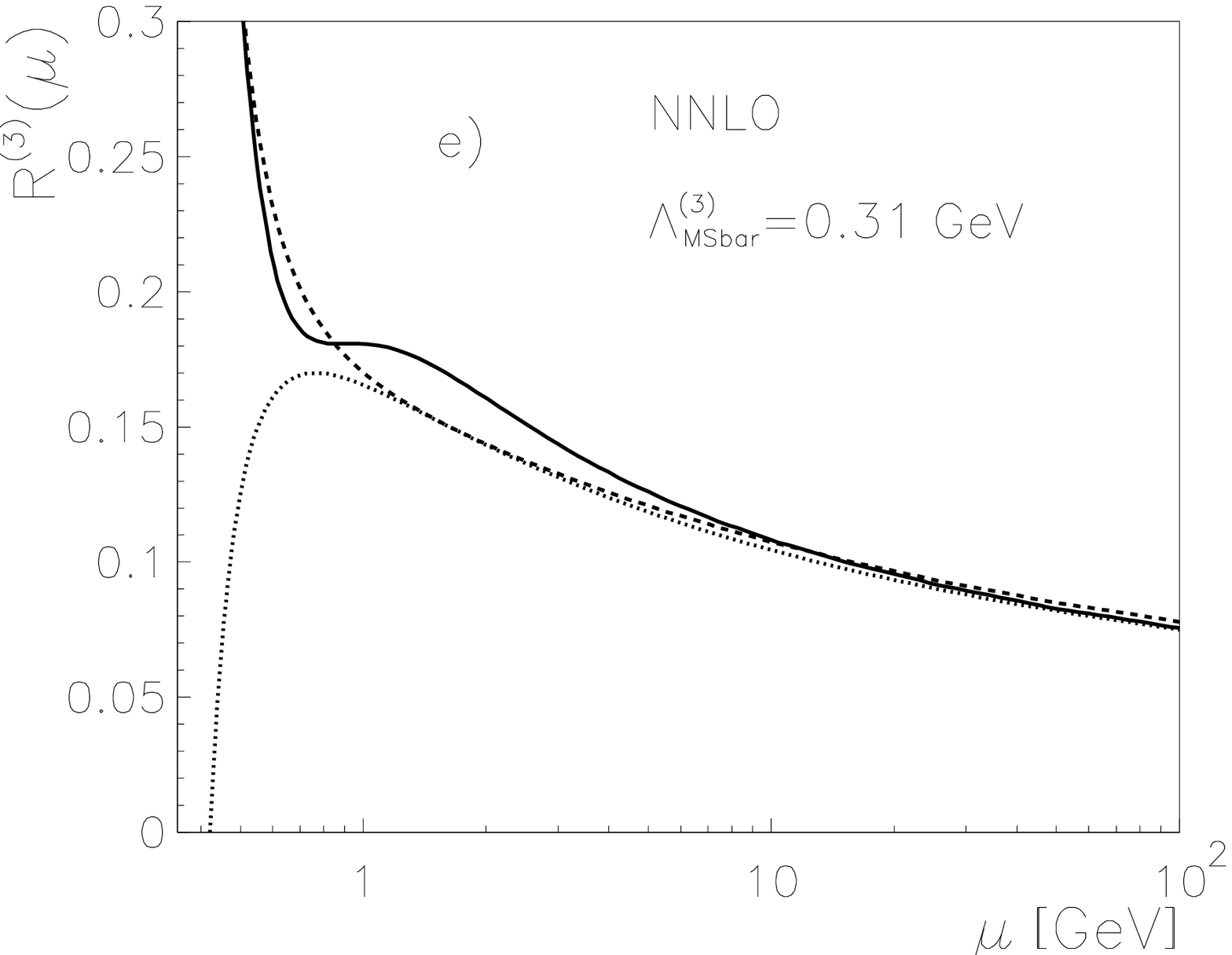,width=8cm}}
\put(0,65){\psfig{file=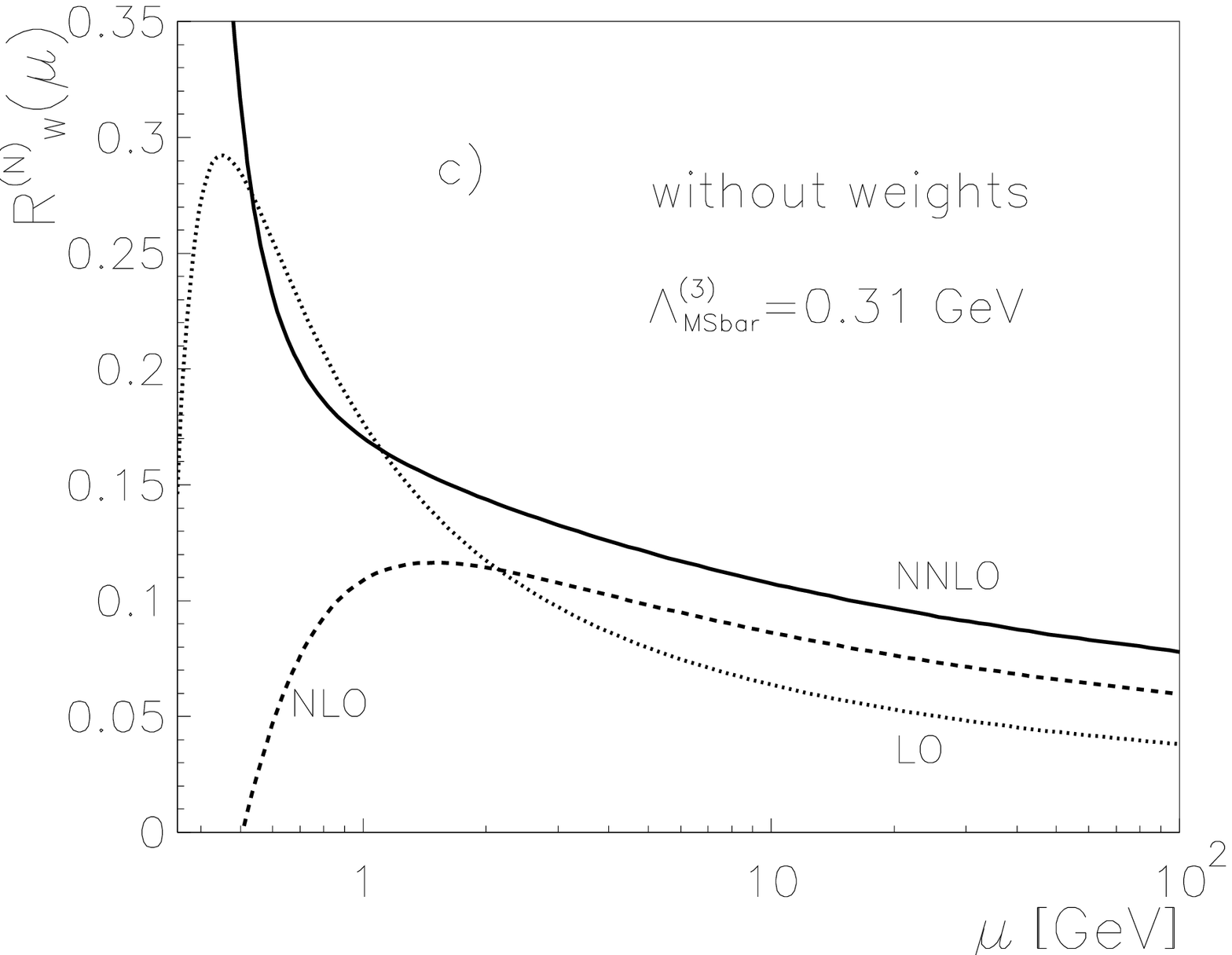,width=8cm}}
\put(80,0){\psfig{file=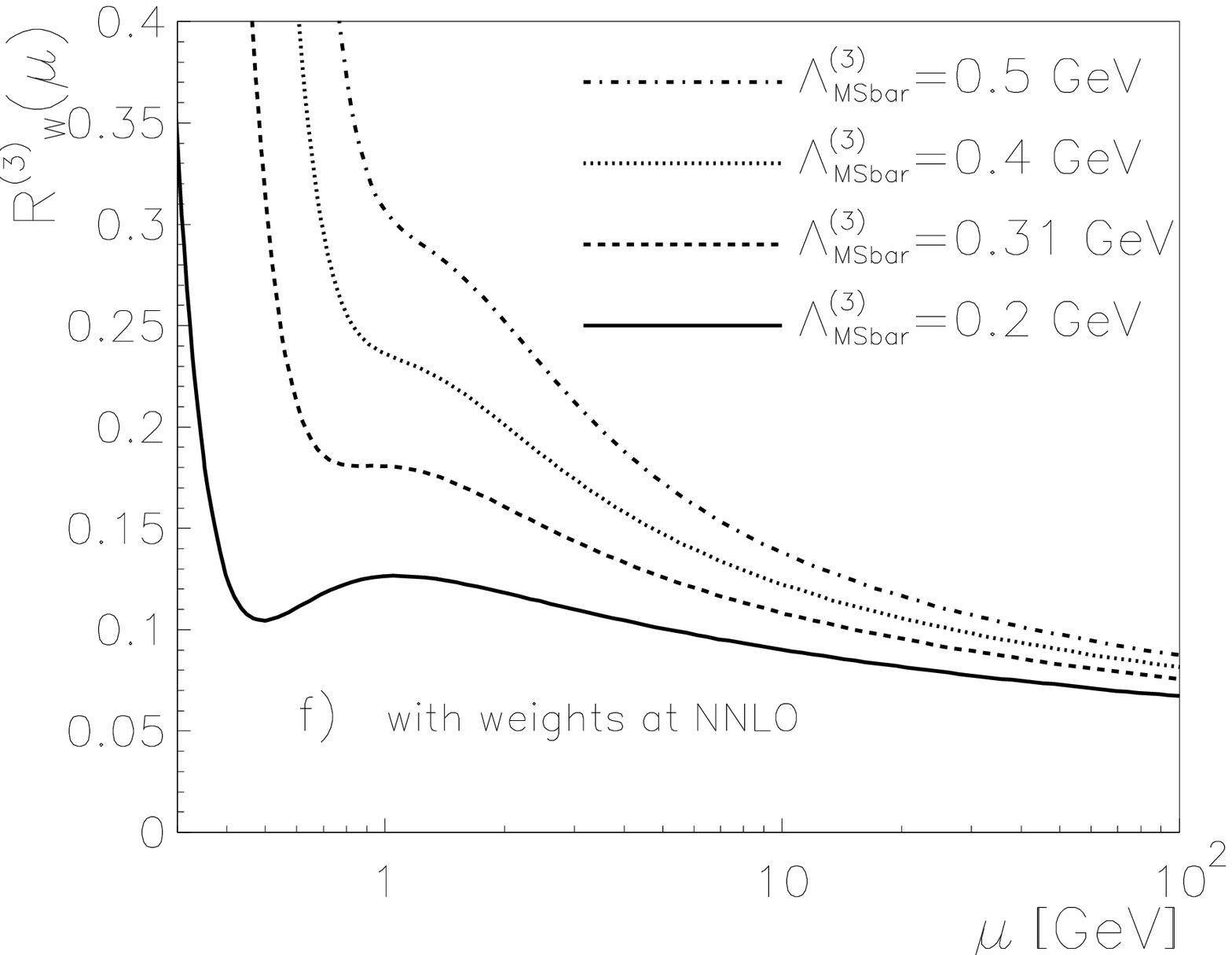,width=8cm}}
\end{picture}
\caption{Scale dependence of  ${\cal R}^{(N)}_{\tau}$ in
the conventional PQCD (a) as well as for the expansion (\ref{Wpert})
in the functions $\widetilde W_n$ (b) and $W_n$ (c).
In d) end e) ${\cal R}^{(2)}_{\tau}$ and ${\cal R}^{(3)}_{\tau}$ of the
conventional PQCD (dotted curves) are compared with the corresponding
approximations using $W_n(a)$ (dashed) and
$\widetilde{W}_n(a)$ (solid). f) the dependence of
${\cal R}^{(3)}_{\tau}$ obtained with functions $\widetilde{W}_n(a)$ on
$\widetilde \Lambda^{(3)}_{\overline{\mathrm{MS}}}$.}
\label{fig1}
\end{figure}

In Figs. \ref{fig1}a-c we compare the scale dependence of the conventional
perturbation expansions of ${\cal R}_{\tau}$ at the LO, NLO and NNLO
with the corresponding expansions in the functions $W_n$ and
$\widetilde W_n$ for
$\widetilde \Lambda^{(3)}_{\overline{\mathrm{MS}}}=0.31$ GeV.
The local maxima of the curves in Fig. \ref{fig1}a define
the PMS choices, the intersections of the NLO and NNLO curves with the LO
one correspond to the ``effective charges''(EC)
approach of \cite{FAC}.
Conventionally the scale $\mu$ is identified with $M_{\tau}$, but this
seemingly natural choice has
a serious drawback as the resulting finite order approximations depend
on the choice of the scheme
\footnote{The point is that in different
schemes the same choice $\mu=M_{\tau}$ leads to different results for
${\cal R}^{(N)}_{\tau}$. Conventionally one works in the
$\overline{\mathrm{MS}}$ RS, but there is no compelling theoretical
argument for this choice. Had we worked, for instance, in MS or MOM schemes
instead, the same choice $\mu=M_{\tau}$ would correspond in
Fig. \ref{fig1}a to the points $\mu_{\mathrm{MS}}=0.68$ GeV and
$\mu_{\mathrm{MOM}}=3.9$ GeV respectively and thus yield
significantly different values of ${\cal R}^{(N)}_{\tau}$. On the other
hand, the scale fixings based on the PMS and EC criteria lead
to the same value of ${\cal R}_{\tau}^{(N)}$ in any scheme.}.

In Fig. \ref{fig1}d-e we compare ${\cal R}^{(N)}_W,~N=2,3$ in the
conventional perturbation theory with the results obtained within
non-power expansion (\ref{Wpert}) for both sets of functions $W_n$ and
$\widetilde{W}_n$. Finally, in Fig. \ref{fig1}f the dependence of
${\cal R}_W^{(3)}$ obtained with the functions $\widetilde{W}_n$
on $\widetilde \Lambda^{(3)}_{\overline{\mathrm{MS}}}$ is displayed.
Several interesting conclusions can be drawn from these figures:
\begin{itemize}
\item
Scale dependence of the NLO and NNLO approximants
${\cal R}_W^{(2)}$ and ${\cal R}_W^{(3)}$ differs, for both $W_n$ and
$\widetilde{W}_n$, significantly from that of the conventional
perturbation theory.
\item There is a
striking difference between the scale dependence of the approximants
${\cal R}_W^{(2)}$ and ${\cal R}_W^{(3)}$, both for $W_n$ and $\widetilde{W}_n$.
\item
There is no region of local stability of ${\cal R}_W^{(3)}$ obtained
with the functions $W_n$, whereas using the functions $\widetilde{W}_n$
there is a plateau for
$\widetilde \Lambda^{(3)}_{\overline{\mathrm{MS}}}\lesssim 0.3$ GeV, but even for
higher values of $\widetilde \Lambda^{(3)}_{\overline{\mathrm{MS}}}$
there is at least a ``knee'' in ${\cal R}_W^{(3)}$.
\item
The value of ${\cal R}_{\tau}^{(3)}$ obtained with functions $\widetilde{W}_n$
is very close to the PMS optimal point of the conventional NNLO approximation.
Remarkably, at this order the approximation obtained with $W_n$ starts to deviate
from the conventional NNLO approximation close to just this stationary point.
\item
The preceding conclusions depend only weakly on the value of
$\widetilde \Lambda^{(3)}_{\overline{\mathrm{MS}}}$ in the reasonably wide
interval $(200,400)$ MeV.
\end{itemize}
%Comprehensive analysis of the quantitative differences between the
%standard perturbative expansions and those based on the functions
%$W_n$ or $\widetilde W_n$ will be presented elsewhere.

Similar analyses of the non-power expansions
introduced by Shirkov \cite{Shirkov} et al. and Cvetic \cite{Cvetic} et al.
could bring interesting new insights.

\vspace*{0.3cm}
\noindent
{\bf Acknowledgments}\\
This work was supported by the Ministry of Education of the
Czech Republic under the project LN00A006 and by the Institutional
Research Project AVOZ1-010-920. Interesting discussions with
E. Gardi and D. Shirkov and the hospitality of the organizers of the RG 2002
Conference are gratefully acknowledged. Two of
us (I.C., J.F.) wish to thank Professor J\"{u}rg Gasser for his kind
hospitality at the Bern University.

\end{document}